\documentclass[prl,aps,twocolumn,superscriptaddress,floatfix,notitlepage,nofootinbib,amssymb,amsmath]{revtex4-1}
\usepackage{mathtools}
\usepackage{epsfig}
\newcommand{\beqn}{\begin{eqnarray}}
\newcommand{\eeqn}{\end{eqnarray}}
\newcommand{\eq}[1]{(\ref{#1})}

\newcommand{\cO}{{\cal O}}
\newcommand{\cS}{{\cal S}}
\newcommand{\cL}{{\cal L}}

\newcommand{\cD}{{\cal D}}

\newcommand{\Tr}{{\mathrm{Tr}\,}}

\newcommand{\lat}{{\mathrm{lat}\,}}

\newcommand{\Cas}{{\mathrm{Cas}}}

\newcommand{\bs}{\boldsymbol}

\newcommand{\avr}[1]{{\left\langle #1 \right\rangle}}
\newcommand{\aavr}[1]{\avr{\!\avr{ #1 }\!}_R}
\def\bbbone{{\mathchoice {\rm 1\mskip-4mu l} {\rm 1\mskip-4mu l} {\rm 1\mskip-4.5mu l} {\rm 1\mskip-5mu l}}}

\begin{document}

\title{
Casimir effect in Yang-Mills theory
}

\author{M. N. Chernodub}
\affiliation{Institut Denis Poisson UMR 7013, Universit\'e de Tours, 37200 France}
\affiliation{Laboratory of Physics of Living Matter, Far Eastern Federal University, Sukhanova 8, Vladivostok, 690950, Russia}
\author{V. A. Goy}
\affiliation{Laboratory of Physics of Living Matter, Far Eastern Federal University, Sukhanova 8, Vladivostok, 690950, Russia}
\author{A. V. Molochkov}
\affiliation{Laboratory of Physics of Living Matter, Far Eastern Federal University, Sukhanova 8, Vladivostok, 690950, Russia}
\author{Ha Huu Nguyen}
\affiliation{Institut Denis Poisson UMR 7013, Universit\'e de Tours, 37200 France}
\affiliation{Faculty of Pedagogy, University of Dalat, Lam Dong 671460, Vietnam}

\begin{abstract}
We study, for the first time, the Casimir effect in non-Abelian gauge theory using first-principle numerical simulations. Working in two spatial dimensions at zero temperature we find that closely spaced perfect chromoelectric conductors attract each other with a small anomalous scaling dimension. At large separation between the conductors, the attraction is exponentially suppressed by a new massive quantity, the Casimir mass, which is surprisingly different from the lowest glueball mass. The apparent emergence of the new massive scale may be a result of the backreaction of the vacuum to the presence of the plates as sufficiently close chromoelectric conductors induce, in a space between them, a smooth crossover transition to a color deconfinement phase. 
\end{abstract}

\date{May 29, 2018}

\maketitle

Quantum fluctuations of virtual particles are affected by the presence of physical objects. This property is a cornerstone of the Casimir effect~\cite{ref:Casimir} which states that the energy of vacuum (``zero-point'') quantum fluctuations should be modified by the presence of physical bodies~\cite{ref:Bogdag,ref:Milton}. The energy shift of the virtual particles has real physical consequences because the Casimir effect leads to appearance a small quantum force, known as the ``Casimir-Polder force''~\cite{Casmir:1947hx}, between neutral objects. The Casimir-Polder forces has been detected in various experiments~\cite{ref:experiment} which support the physical significance of the vacuum energy of virtual particles. 

Generally, a theoretical calculation of the Casimir-Polder forces is a difficult analytical problem even in noninteracting field theories since the energy spectrum of vacuum fluctuations, apart from simplest geometries, cannot be accurately determined. Therefore the Casimir effect is often studied using certain analytical approximations such as the proximity-force calculations~\cite{ref:proximity} and various numerical tools~\cite{Johnson:2010ug} which includes worldline approaches~\cite{Gies:2006cq} and methods of lattice field theories~\cite{ref:Oleg,ref:paper:1,ref:paper:2}.

In interacting (gauge) theories the calculations become even more involved. In a phenomenologically relevant case of quantum electrodynamics a correction to the Casimir-Polder force coming from fermionic vacuum loops is given by the second order perturbation theory, which turns out to be negligibly small due to the weakness of the electromagnetic coupling~\cite{Bordag:1983zk,ref:Milton}. Perturbative calculations in finite-volume geometries of non-Abelian gauge theories were also addressed~\cite{ref:perturbative:YM}. 

In strongly coupled theories the interactions may not only lead so a noticeable modification of the Casimir-Polder forces, but they may also affect the structure of the vacuum itself. For example, the Casimir effect in between two parallel plates leads to strengthening of a chiral finite-temperature phase transition in a four-fermion effective field theory~\cite{Flachi:2013bc}. The presence of the boundaries effectively restore the chiral symmetry in an otherwise chirally broken phase both in plane~\cite{Tiburzi:2013vza} and in cylindrical~\cite{Chernodub:2016kxh} geometries, revealing that the chiral properties of the system depend on the geometry of the system. The interactions may even change the overall sign of the Casimir--Polder force in certain fermionic systems with condensates~\cite{Flachi:2017cdo} and in the $\mathbb{C}P^{N-1}$ model on an interval~\cite{Flachi:2017xat,Betti:2017zcm}. First-principle numerical simulations show that the presence of the boundaries affects also non-perturbative (de)confining properties of certain bosonic gauge systems~\cite{ref:paper:1,ref:paper:2}. 

In our paper we initiate an investigation of the Casimir effect in Yang-Mills theory which has inherently nonperturbative vacuum structure. In the absence of matter fields the zero-temperature Yang-Mills theory exhibits two interesting phenomena, mass gap generation and color confinement, which influence the excitation spectrum of its phenomenologically relevant counterpart in particle physics, Quantum Chromodynamics. We use first-principle numerical methods to determine the Casimir-Polder forces in finite geometries of the non-Abelian vacuum and, inversely, the influence of the finite-geometry on the non-Abelian structure of the theory. 

In order to simplify our analysis we consider a zero-temperature Yang-Mills theory in (2+1) spacetime dimensions which exhibits both mass gap generation and color confinement similarly to the theory in 3+1 dimensions. In 2+1 dimensions the boundary conditions for the (gauge) fields are usually formulated at one-dimensional manifolds rather then at surfaces that are more relevant to theories with three spatial dimensions. To be precise, we concentrate on a simplest geometry given by two parallel static wires along the $x_2$ direction separated by a finite distance $R$ along the $x_1$ axis.

A simplest version of the Casimir effect in a gauge system can be formulated in a $U(1)$ Maxwellian gauge theory with the Lagrangian 
\beqn
\cL_{U(1)} = - \frac{1}{4} f_{\mu\nu} f^{\mu\nu}, \qquad f_{\mu\nu} = \partial_\mu a_\nu - \partial_\nu a_\mu\,,
\label{eq:U1:L}
\eeqn
where $a_\mu$ is an Abelian gauge field. Restricting ourselves to ideal situations, one may impose either boundary conditions corresponding to a material made of a perfect electric conductor (with normal magnetic and tangential electric components vanishing at conductor's boundary) or its dual analogue, an ideal magnetic conductor (in which magnetic and electric components exchange their roles). In our article we consider boundaries made of electric-type boundary conditions, which, in two spatial dimensions, are given by the following local condition:
\beqn
\epsilon^{\mu\alpha\beta} n_\mu(x) f_{\alpha\beta} (x) = 0\,,
\label{eq:U1:BC}
\eeqn
where $n_\mu(x)$ is a vector normal to the boundary at the point~$x$. In the geometry of two parallel wires in two spatial dimensions the vacuum fluctuation of the $U(1)$ gauge field lead to the attractive potential between the wires, $V_\Cas(R) = - \zeta(3)/(16 \pi R^2)$~\cite{Ambjorn:1981xw,ref:paper:1}, where $\zeta(x)$ is the zeta function.

The Lagrangian of Yang-Mills theory has the form:
\beqn
\cL_{YM} = - \frac{1}{4} F_{\mu\nu}^a F^{\mu\nu,a},
\label{eq:L:YM}
\eeqn
where $F^a_{\mu\nu} = \partial_\mu A_\nu^a - \partial_\nu A_\mu^a + g f^{abc} A_\mu^b A_\nu^c$ is the field-strength tensor of the non-Abelian (gluon) field $A_\mu^a$ with $a = 1, \dots N_c^2 -1$, and $f^{abc}$ are the structure constants of the $SU(N_c)$ gauge group. A non-Abelian analogue of the perfect conductor condition~\eq{eq:U1:BC} is straightforward:
\beqn
\epsilon^{\mu\alpha\beta} n_\mu(x) F^a_{\alpha\beta} (x) = 0\,, \qquad a = 1, \dots, N_c^2 -1\,.
\label{eq:YM:BC}
\eeqn
In our geometry these perfectly conducting chromoelectric wires are positioned at the straight lines $x_1 = 0$ and $x_1 = R$, so that the normal vector is $n_\mu = \delta_{\mu1}$.

In a tree order one may formally set $g=0$ so that both the Yang-Mills theory~\eq{eq:L:YM} and the boundary conditions~\eq{eq:YM:BC} are reduced to $N_c^2 - 1$ noninteracting copies of the Maxwell electrodynamics~\eq{eq:U1:L} with the $U(1)$ boundary conditions~\eq{eq:U1:BC}. Thus in a tree order all $N_c^2 - 1$ gluons contribute additively to the Casimir energy density:
\beqn
V_\Cas^{\mathrm{tree}} = - (N^2_c - 1) \frac{\zeta(3)}{16 \pi R^2}\,.
\label{eq:V:Cas:tree}
\eeqn
However, the non-Abelian theory is an essentially nonperturbative system and therefore the Casimir potential of interacting gluons may substantially differ from the naive tree level expression~\eq{eq:V:Cas:tree}. In order to elucidate this question we use the first-principle lattice simulations.

The lattice version of the $N_c = 2$ Yang-Mills theory \eq{eq:L:YM} is given in terms of the $SU(2)$ link variables $U_l$ residing on the links $l \equiv l_{x,\mu}$ of the Euclidean cubic lattice $L_s^3$ with periodic boundary conditions in all three directions. The path integral is given by the integration with the Haar measure over all link variables $U_l$:
\beqn
Z = \int \cD U \, e^{- S[U]},
\label{eq:Z}
\eeqn
where we use the standard plaquette action
\beqn
S[U] \equiv \sum_P S_P = \sum_P \beta_P \Bigl\{ 1 - \frac{1}{2}  \Tr U_P\Bigr\}\,,
\label{eq:S}
\eeqn
with the plaquette fields $U_{P_{x,\mu\nu}} = U_{x,\mu}U_{x+\hat\mu,\nu}U^\dagger_{x+\hat\nu,\mu} U^\dagger_\nu$, the notation $\hat\mu$ denotes a unit lattice vector in the positive $\mu$ direction. In the absence of the Casimir wires the lattice coupling constants $\beta_P$ are independent of plaquettes, $\beta \equiv \beta_P$, where the bulk coupling constant
\beqn
\beta = \frac{4}{a g^2}\,,
\label{eq:beta}
\eeqn
is related to the lattice spacing $a$. The quantity $g^2$, which has the dimension of mass, becomes the physical coupling of the continuum Yang-Mills theory in the limit $a \to 0$. The lattice $U_l$ and continuum $A_\mu^a$ fields are related as follows: $U_{l_{x,\mu}} = P e^{ i g \int_{x}^{x+a\hat\mu} d x^\nu {\hat A}_\nu(x)}\simeq e^{i a g {\hat A}_\mu(x)}$ with ${\hat A}_\mu = T^a A_\mu^a$, and $T^a$ are the generators of the non-Abelian gauge group, $[T^a,T^b]= 2 i f^{abc} T^c$.

The lattice version of the chromoelectric boundaries~\eq{eq:YM:BC} is enforced by the space-dependent coupling~\cite{ref:paper:1}:
\beqn
\beta_P = \left\{
\begin{array}{lll}
\beta, & \quad & P \notin \cS\,, \\
\lambda_w\beta, & \quad & P \in \cS\,, \\
\end{array}
\right.
\label{eq:beta:P}
\eeqn
where $\cS = \cS_1 \cup \cS_2$ is a union of the worldsurfaces of the wires. In our geometry Eq.~\eq{eq:beta:P} implies that $\beta_P = \lambda_w\beta$ at the plaquettes $P_{x,23}$ with $x_1=0$ and $x_1 = R$, and $\beta_P = \beta$ otherwise. Then we increase the lattice coupling at the wires, $\lambda_w \to +\infty$, so that the tangential (to each wire) component of the chromoelectric field vanishes ($U_{P_{23}} \to \bbbone$ in lattice terms), leading to the perfect ``chromometallic'' conditions~\eq{eq:YM:BC}.

The energy of vacuum fluctuations of gluon field is related to a local expectation value of its energy density,
\beqn
T^{00} = \frac{1}{2} \left({\bs B}_z^2 + {\bs E}_x^2 + {\bs E}_y^2 \right),
\label{eq:T00:Minkowski}
\eeqn
which is a component of the energy-momentum tensor associated with the Yang-Mills Lagrangian~\eq{eq:L:YM}:
\beqn
T^{\mu\nu} = - F^{\mu\alpha,a} F^{\nu,a}_{\ \alpha} + \frac{1}{4} \eta^{\mu\nu} F_{\alpha\beta}^a F^{\alpha\beta,a}.
\eeqn
In a Minkowski spacetime with the metric $(+,-,-)$ one has $F^a_{01} = E^a_x$,  $F^a_{02} = E^a_y$ and $F^a_{12} = - B^a_z$ with $a=1,2,3$, and ${\bs E}_x^2 \equiv (E^a_x)^2$ etc.

After a Wick rotation to a Euclidean space the energy density~\eq{eq:T00:Minkowski} transforms to
\beqn
T^{00}_E = \frac{1}{2} \left({\bs B}_z^2 - {\bs E}_x^2 - {\bs E}_y^2 \right).
\label{eq:T00:Euclidean}
\eeqn
The Euclidean lattice corresponding to a zero temperature theory is symmetric under $\pm \pi/2$ rotations about the $x \equiv x_1$ axis, which imposes the equivalence of the fluctuations of the tangential gauge fields, $\avr{{\bs B}_z^2} = \avr{{\bs E}_x^2}$. Thus the expectation values of first two terms in Eq.~\eq{eq:T00:Euclidean} cancel each other in the normalized energy density:
\beqn
{\cal E}_R(x) & 
= & \avr{T^{00}_E(x)}_{R} - \avr{T^{00}_E(x)}_{0} \nonumber \\
& \equiv & \frac{1}{2} \left(\avr{{\bs E}_y^2}_0 - \avr{{\bs E}_y^2(x)}_R \right)\,,
\label{eq:E:norm}
\eeqn
where the subscripts ``0''  and ``$R$''  indicate that the expectation value is taken, respectively, in the absence of the wires and in the presence of the wires separated by the distance $R$. Due to the normalization the ultraviolet divergencies cancel in Eq.~\eq{eq:E:norm} so that ${\cal E}_R(x)$ provides us with a local finite quantity, the Casimir energy density, which is equal to a change in the energy density of the vacuum fluctuations due to the presence of the wires. 

In our geometry the energy density~\eq{eq:E:norm} depends only on the coordinate $x_1$ normal to the wires. Therefore it is natural to introduce the (Casimir) energy density per unit length of the wires:
\beqn
V_{\Cas}(R) = \int d x_1\, {\cal E}_R(x_1)\,.
\label{eq:V:Cas}
\eeqn
In the lattice regularization the Casimir energy density, given by Eqs.~\eq{eq:E:norm} and \eq{eq:V:Cas}, takes the following form:
\beqn
V^{\lat}_{\Cas}(R) = - \aavr{S_{P_{23}}}^{\lat},
\label{eq:V:Cas:lat}
\eeqn
where the plaquette $P_{23}$ is oriented along wires' direction ($\mu=2$) and the Euclidean time ($\nu=3$), and 
\beqn
\avr{\!\avr{\cO(x)}\!}_R^{\lat} = \sum_{x_1 = 0}^{L_s - 1} \left[ \avr{\cO(x_1)}_R - \avr{\cO}_0 \right]\,,
\label{eq:avr:O:lat}
\eeqn
is a lattice expression for the excess of the expectation value of the operator $\cO$ evaluated per a unit wire length. Thus the lattice expression for the Casimir energy density~\eq{eq:V:Cas:lat} is given by a normalized expectation value of the components of the plaquette lattice action~\eq{eq:S}, that are parallel to the Euclidean worldsheets of the wires.

The physical density of the Casimir energy at a physical distance $R = a R_\lat$ between the wires is determined by the simple scaling formula $V^{\mathrm{phys}}_{\Cas}(R) = a^{-2} V^{\lat}_{\Cas}(R/a)$ where $a$ is the lattice spacing in physical units. In $SU(2)$ gauge theory the lattice gauge coupling $\beta$ is related to the lattice spacing $a$ according to Eq.~\eq{eq:beta} in which the massive parameter $g^2$ becomes the continuum coupling squared in the continuum limit $a \to 0$. Ideally, the continuum physics is reached as $\beta \to \infty$ while in practice one deals with finite values of the lattice coupling $\beta$ which affect the extrapolation to continuum with $O(a^n)$ corrections. We estimate that finite size effects may lead to numerically significant (of the order of 10-15\%) corrections to the non-Abelian Casimir energy.

We improve the continuum scaling at finite $\beta$ in three steps. Firstly, in order to reduce the finite-size corrections it is advantageous to use the mean-field improved coupling which is expressed via the average plaquette~\cite{ref:beta:imp,Teper:1998te}:
\beqn
\beta_{\mathrm{I}}(\beta) = \beta \cdot \frac{1}{2} \avr{\Tr U_P}(\beta)\,.
\label{eq:beta:imp}
\eeqn
Secondly, we express the physical lattice spacing $a$ via the phenomenologically determined series over $1/\beta_I$~\cite{Teper:1998te}
\beqn
a \sqrt{\sigma} = \frac{1.341(7)}{\beta_I}  - \frac{0.421(51)}{\beta_I^2} +  O\bigl(1/\beta^3_I\bigr)\,,
\label{eq:a:sigma}
\eeqn
where $\sigma$ is the tension of the confining (fundamental) Yang-Mills string at zero temperature. In the selected range of the coupling constant $\beta$ the higher-order terms in Eq.~\eq{eq:a:sigma} are numerically irrelevant. 

Thirdly, we notice that in the lattice perturbation theory the expectation value the lattice plaquette operator $\avr{\Tr U_{\mu\nu}}$ acquires radiative corrections, both of additive and multiplicative nature. The additive corrections -- which correspond to the UV-divergent perturbative vacuum contributions --  are automatically removed from the Casimir energy by the subtraction scheme~\eq{eq:avr:O:lat}. The multiplicative correction originates from the fact that the physically relevant quantity is the product $\beta^4 \avr{\Tr U_{\mu\nu}} \sim a^{-4} \avr{\Tr U_{\mu\nu}} \sim \avr{F_{\mu\nu}^2}_{\mathrm{phys}}$ and not the expectation value of the plaquette itself (for example, it is the former quantity that determines the physical value of the nonperturbative gluon condensate in the Yang-Mills theory~\cite{Hietanen:2004ew}). In order to improve the finite-size scaling of our results we thus rescale the expectation value of the plaquette operator with the improved value of the coupling constant~\eq{eq:beta:imp}: $\avr{\Tr U_{\mu\nu}} \to \avr{\Tr U_{\mu\nu}}_I = (\beta_I/\beta)^4 \avr{\Tr U_{\mu\nu}}$. As we will see shortly below, this phenomenological procedure -- which substitutes the multiplicative factor $\beta^4$ by its improved version $\beta_I^4$ -- leads to a nearly perfect scaling of the Casimir energy at intermediate values of the lattice coupling~$\beta$. 

Summarizing, the scale-improved relation for the Casimir energy density in the continuum limit, expressed via numerically calculable quantities on the lattice is given by the following formula,
\beqn
V_{\Cas}(R) = - \frac{1}{\sigma} \frac{1}{a^2(\sigma,\beta)} \left(\frac{\beta_I}{\beta}\right)^4 \avr{\!\avr{S_{P_{23}}}\!}_{R/a}^{\lat}, \quad
\label{eq:V:Cas:ph}
\eeqn
where the lattice spacing $a = a(\sigma,\beta)$ and the mean-field improved lattice coupling $\beta_I$ are given in Eqs.~\eq{eq:a:sigma} and \eq{eq:beta:imp}, respectively. The wires are separated by the physical distance $R = R_\lat a$. In the weak coupling limit the expression~\eq{eq:V:Cas:ph} tends to its natural form since in this limit $\beta_{I}(\beta) \to \beta$ and $\avr{\Tr U_{\mu\nu}}_{I} \to \avr{\Tr U_{\mu\nu}}$, and we recover from Eq.~\eq{eq:V:Cas:ph} the naive expression~\eq{eq:V:Cas:lat}.

In our numerical simulations we use methods successfully adopted for studies of the Casimir forces in Abelian gauge theories in Refs.~\cite{ref:paper:1,ref:paper:2,ref:paper:3}. We generate gauge-field configurations using a Hybrid Monte Carlo algorithm which combines standard Monte-Carlo methods~\cite{ref:20} with the molecular dynamics approach. The latter incorporates a second-order minimum norm integrator~\cite{ref:21} with several time scales~\cite{ref:22}, that allows us to equilibrate the integration errors accumulated at and outside worldsheets of the wires at which the Casimir boundary conditions are imposed. Long autocorrelation lengths in Markov chains are eliminated by overrelaxation steps (5 steps between trajectories) which separate gauge field configurations far from each other~\cite{ref:20}. We works on the $32^3$ lattice. For average procedure we use 250000 trajectories.

The non-Abelian permittivity of the wires is fixed by the strength of the coupling constant $\lambda_w\beta$ at the Euclidean worldsurfaces of the wires. At a large wire coupling $\lambda_w\beta$ the wires behave as almost ideal conductors which force all tangent components of the chromoelectric field to vanish at each point of the wire, $F_\|^a \to 0$.

\begin{figure}[!thb]
\begin{center}
\includegraphics[scale=0.5,clip=true]{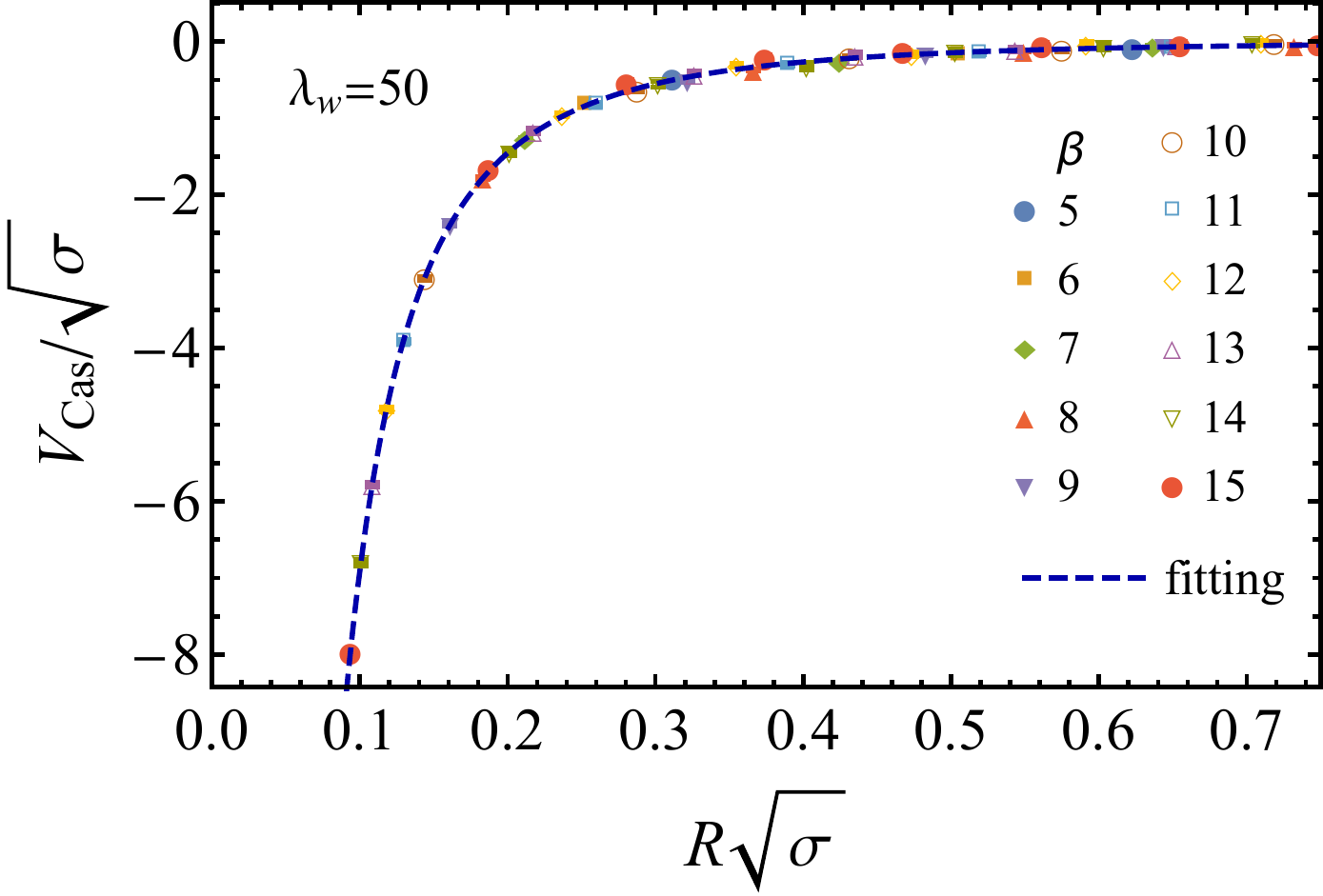}
\end{center}
\vskip -4mm 
\caption{The Casimir potential $V_{\Cas}$ as the function of the distance $R$ between the wires in units of the string tension~$\sigma$. The potential is given for the large excess of wire's coupling ($\lambda_w = 50$) and at various bulk lattice couplings~$\beta$.}
\label{fig:V}
\end{figure}

In Fig.~\ref{fig:V} we show the non-Abelian Casimir energy~\eq{eq:V:Cas:ph} as the function of the distance between the wires $R$ for various values of the bulk lattice coupling $\beta$ and the fixed excess  $\lambda_w = 50$ of the coupling constant at the Euclidean worldsurfaces of the wires.

In order to show a good physical scaling of the Casimir potential, we plot in Fig.~\ref{fig:V} the potential at various lattice coupling constants $\beta$ corresponding to different values of the physical lattice spacing $a$. We obtain an excellent scaling of our results because the Casimir potential obtained at different values of $\beta$ match a single curve as the function of interwire distance $R$, both plotted in terms of the physical string tension $\sigma$. We get also similar physical scaling for other values of $\lambda_w$.

We fit the Casimir potential by the following function:
\beqn
V^{\mathrm{fit}}_{\Cas}(R) = 3 \frac{\zeta(3)}{16 \pi} \frac{1}{R^\nu \sigma^{(\nu-2)/2}} e^{- M_{\Cas} R},
\label{eq:V:fit}
\eeqn
where $\nu$ and $M_C$ are the free parameters determined from the best fit. The power of $\sigma$ in denominator in Eq.~\eq{eq:V:fit} is chosen to keep the correct dimension [mass${}^2$] of the Casimir potential as it corresponds to the Casimir energy of the non-Abelian fluctuations between the wires calculated per unit length of the wire. 

The fitting function~\eq{eq:V:fit} has a transparent physical meaning. The exponent $\nu$ in the fitting function~\eq{eq:V:fit} corresponds to an eventual anomalous dimension of the Casimir potential at short distances. The quantity $M_\Cas$, which we call the Casimir mass, corresponds to an effective screening of the Casimir potential at large distances due to nonperturbative mass gap generation in the non-Abelian gauge theory. In the absence of interactions the mass gap is absent, $M_\Cas = 0$, while the anomalous dimension is equal to its canonical value, $\nu = 2$, so that the phenomenological potential~\eq{eq:V:fit} is naturally reduced to its tree-level expression~\eq{eq:V:Cas:tree}.

\begin{figure}[!thb]
\begin{center}
\includegraphics[scale=0.5,clip=true]{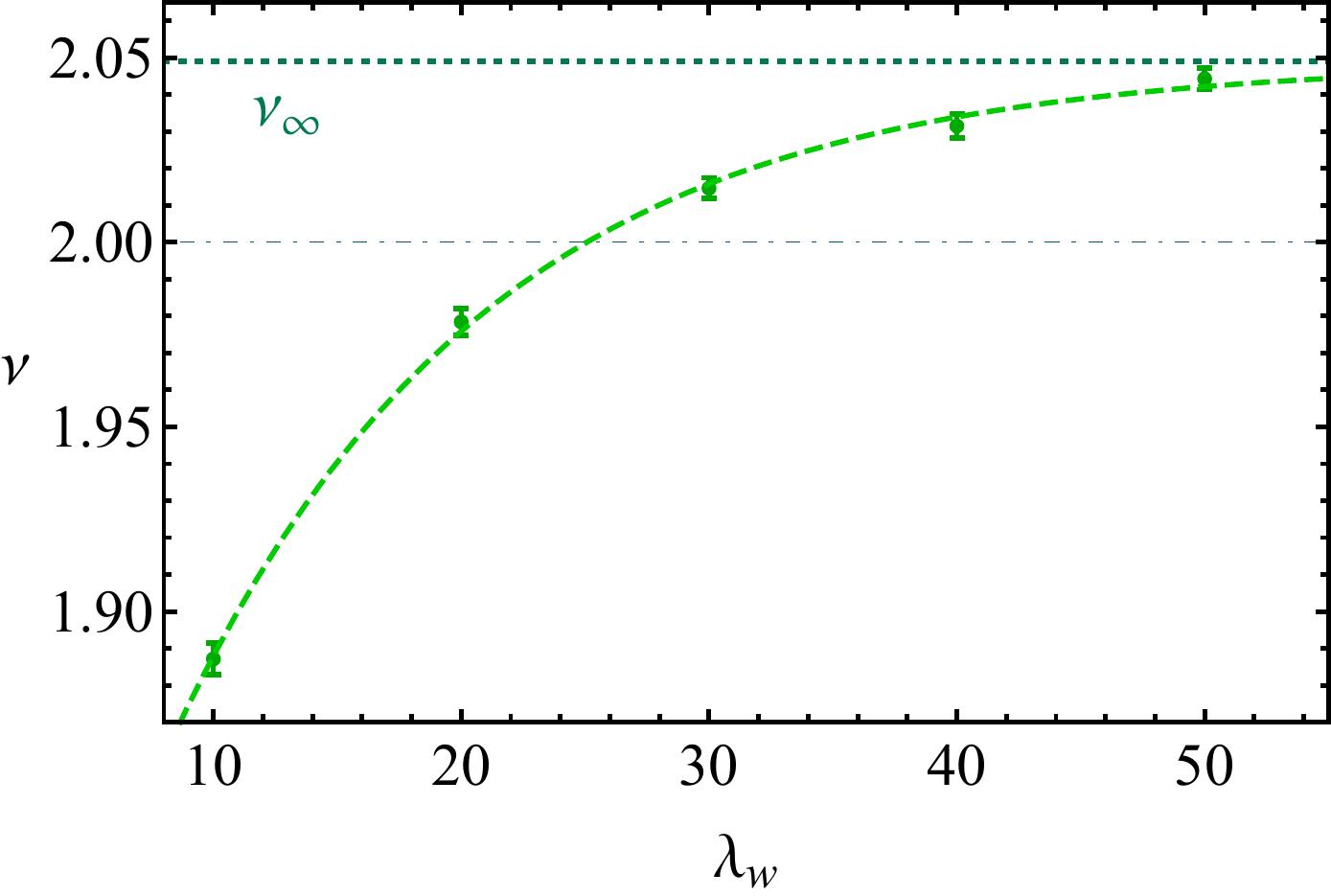}
\end{center}
\vskip -4mm 
\caption{The power $\nu$ of the Casimir potential as the function of the wire excess strength $\lambda_w$. The dashed line represents the best asymptotic fit~\eq{eq:O:fit}, the thick short-dashed line corresponds to the asymptotic value~\eq{eq:nu:infty}, and the thin dot-dashed line denotes the power $\nu=2$ of the noninteracting theory.}
\label{fig:nu}
\end{figure}

\begin{figure}[!thb]
\begin{center}
\includegraphics[scale=0.5,clip=true]{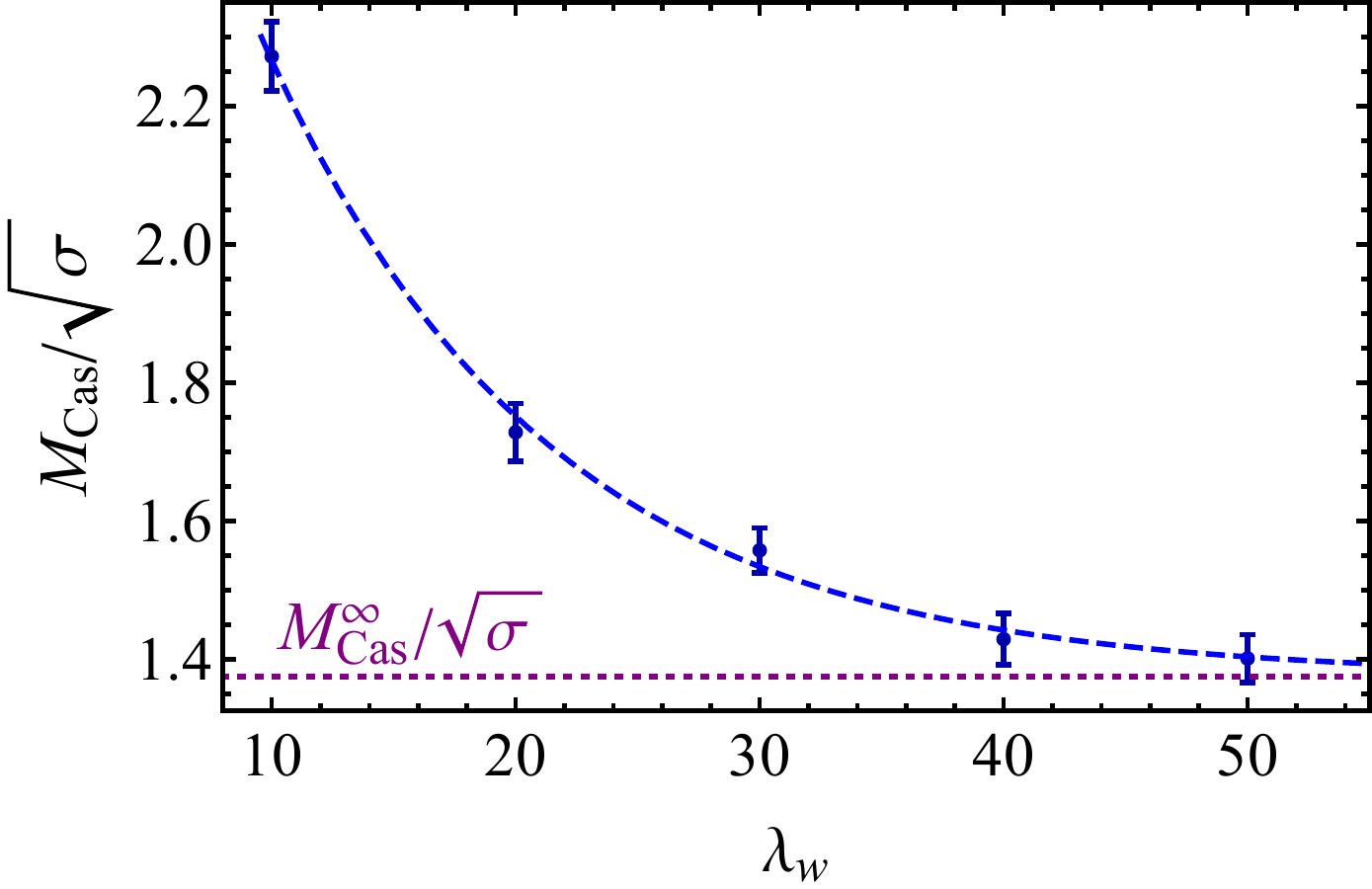}
\end{center}
\vskip -4mm 
\caption{The Casimir mass $M_\Cas$ as the function of the strength of the wire $\lambda_w$. The thick dashed lines denotes the best fit by function~\eq{eq:O:fit}, and the short-dashed horizontal line denotes the asymptotic value~\eq{eq:M:infty}.}
\label{fig:M}
\end{figure}

The best fit of the the Casimir potential with the almost-perfect wires ($\lambda_w = 50$) is shown in Fig.~\ref{fig:V} by the dashed line. The dependences of the best-fit values of the power $\nu$ and the Casimir mass $M_\Cas$ on the strength of the wire $\lambda_w$ are shown in Figs.~\ref{fig:nu} and \ref{fig:M}, respectively. We found that the numerical data for the power and the Casimir mass can be fitted by a simple exponential fit:
\beqn
\cO(\lambda_w) = \cO^\infty + \alpha_\cO e^{- \lambda_w/\lambda_w^\cO},
\label{eq:O:fit}
\eeqn
where $\cO = \nu, M_\Cas$, and $\cO^\infty$, $\alpha_\cO$ and $\lambda_w^\cO$ are the fitting parameters. It turns out that the quantities $\nu$ and $M_\Cas$ rapidly approach, with $\lambda_w^\nu \simeq \lambda_w^{M_\Cas} = 12(1)$, the corresponding asymptotic values $\cO^\infty \equiv \lim_{\lambda_w \to \infty}\cO(\lambda_w)$ in the perfect-wire limit.

According to the fit in Fig.~\ref{fig:nu} the power $\nu$ slightly overshoots the standard free-field value $\nu=2$:
\beqn
\nu_\infty = 2.05(2),
\label{eq:nu:infty}
\eeqn
implying that the self-interaction of the gluon fields may lead to a small anomalous scaling dimension of the Casimir potential.

The asymptotic value of the Casimir mass $M_\Cas$, corresponding to the energy of the vacuum fluctuations of non-Abelian gauge field between perfect wires ($\lambda_w \to \infty$), 
\beqn
M^\infty_{\Cas} = 1.38(3) \sqrt{\sigma}\,,
\label{eq:M:infty}
\eeqn
is shown by the horizontal dashed line in Fig.~\ref{fig:M}. Surprisingly, the Casimir mass~$M_\Cas$ turns out to be much smaller than the mass 
\beqn
M_{0^{++}} \approx 4.7 \sqrt{\sigma}\,,
\label{eq:M:glueball}
\eeqn
of the lowest $0^{++}$ glueball in SU(2) gauge theory. The latter was calculated in lattice simulations of Refs.~\cite{Teper:1998te,Athenodorou:2016ebg}. According to Fig.~\ref{fig:M} the enhancement of the strength of the wires leads to a diminishing of the Casimir mass. On the contrary, as the wire weakens, the Casimir mass moves towards the higher masses so that we may expect that in the weak-wire limit $\lambda_w \to 1$ the Casimir mass $M_\Cas$ may naturally approach the mass  of the lightest glueball, $M_{0^{++}}$, although the simple form of the fitting function~\eq{eq:O:fit} does not allow us to make this conclusion more precise at the present stage.

One may also suggest that in the volume between finitely-separated parallel wires the gluons behave as if they are subjected to a heat bath at finite temperature. In a Euclidean formulation of a finite-temperature theory in a thermodynamic equilibrium the temporal direction is compactified to a circle with the length $1/T$, making the fields periodic along this direction.

Contrary to the case of finite temperature, the periodicity of the gluon fields is evidently absent in the Casimir setup. However, similarity to the $T \neq 0$ case, the perfectly conducting wires do indeed restrict allowed frequencies of free gluons with certain polarizations. In particular, the propagator of free gluons in the Feynman gauge corresponds to the Neumann boundary condition for the normal (with respect to the boundary) gluon component $A^a_\perp \equiv A^a_1$~\cite{Peterson:1981yy}. In the limit of a small-separation between the plates the Neumann boundaries dimensionally reduce the dynamics of the normal gluon components $A^a_\perp$ to (1+1) dimensional spacetime with the tangential coordinate $x_\| = (x_2,x_0\equiv x_3)$. Thus the normal gluon component $A^a_\perp$ in the Casimir setup plays a role of the timelike gluon $A^a_0$ at $T \neq 0$. In finite-temperature (2+1) dimensional gauge theories the $A^a_0$ gluons are spatially correlated with the screening mass typically proportional to $g \sqrt{T}$ (see, for example, \cite{Alves:2002tg}). Therefore in our case we may expect that the normal gluon components $A^a_\perp$ are correlated along the conducting wires with the ``Casimir'' screening mass $M_{g,\Cas}^2 = c_D g^2/(2 \pi R)$, where $c_D$ is a constant of the order of unity. 

Yang-Mills theories are known to experience a deconfinement phase transition at sufficiently high temperature. In particular, in (2+1) dimensions the critical temperature of the transition has been determined in Ref.~\cite{Teper:1993gp}. Given the mentioned analogy, one may expect  the gluonic vacuum in between sufficiently close wires may enter a deconfinement-like regime. A similar conclusion may also be drawn from properties of a confining compact QED in finite geometries~\cite{ref:paper:3}. In order to check this idea we calculate numerically the deconfinement order parameter, the Polyakov line $L$, which has a vanishing expectation value in the confinement phase, $\avr{L} = 0$,  and a nonzero value in the deconfinement phase, $\avr{L} \neq 0$. 

In the lattice formulation of Yang-Mills theory the Polyakov line is given by an ordered product of the non-Abelian matrices along the temporal ($\mu=3$) direction:
\beqn
L_{\bs x} = \frac{1}{2} \Tr \prod_{x_3=0}^{L-1} U_{{\bs x},x_3;3},
\label{eq:L}
\eeqn
where ${\bs x} \equiv (x_1,x_2)$ is the spatial two-dimensional coordinate. Notice that the Polyakov line~\eq{eq:L} is defined along the long temporal direction while in the finite temperature theory the line is directed along the short compactified time. This property shows a difference between the Casimir-like geometry and a finite-temperature theory. 

\begin{figure}[!thb]
\begin{center}
\includegraphics[scale=0.65,clip=true]{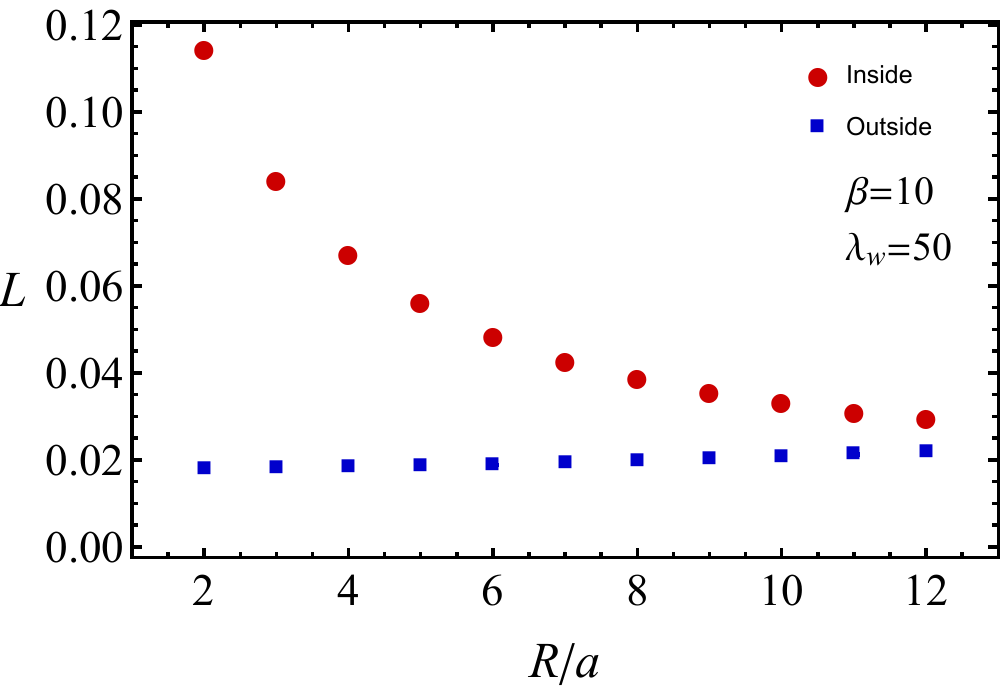}
\end{center}
\vskip -4mm 
\caption{The expectation value of the Polyakov line in between and outside the wires for the bulk gauge coupling $\beta=10$ and the strong in-wire coupling $\lambda_w = 50$.}
\label{fig:L}
\end{figure}

We calculated the Polyakov line separately in the volume inside and outside the wires:
\beqn
L_{\ell} = \frac{1}{{S_\ell}} \biggl\langle\biggl|\, \sum_{{\bs x} \in {S_\ell}} L_{\bs x}\, \biggr| \biggr\rangle,
\label{eq:P:in:out}
\eeqn
where $S_a$ is the two-dimensional area in between ($\ell =\mathrm{in}$) and outside ($\ell =\mathrm{out}$) the wires. In Fig.~\ref{fig:L} we show the expectation values of the Polyakov line as a function of the distance between the wires. One can clearly see that as the distance between the wires gets smaller the expectation value of the Polyakov line in between the wires increases thus signaling an approach to a deconfinement regime. However, we have not found a critical behavior of the Polyakov lines, in agreement with the smooth behavior of the Casimir potential as a function of the interwire distance~$R$, Fig.~\ref{fig:V}. The Polyakov line outside the wires is largely insensitive to the separation~$R$. We conclude that the gluon dynamics in between the wires becomes deconfining, while the transition in between the confining and deconfining regions is, most probably, a smooth non-critical crossover. The latter conclusion is supported by a finite-volume nature of the Casimir effect.

Summarizing, we studied for a first time the Casimir effect in a zero-temperature non-Abelian gauge field theory. As a simplest example we considered SU(2) gauge theory in low $2+1$ dimensions, in which vacuum fluctuations of non-Abelian gauge fields are restricted by the presence of two parallel straight wires separated by a finite distance~$R$. The wires act as perfectly {\emph{chromo}}-conducting boundaries so that all non-Abelian components of the chromoelectric field tangent to the direction of the wires vanish at each point of every wire.

We have found that at large distances between the wires the attractive Casimir interaction is an exponentially diminishing function of the interwire separation. This effect is not an unexpected phenomenon given the existence of the mass gap generation phenomenon in the zero-temperature Yang-Mills theory. However, the infrared suppression of the Casimir interaction between the wires is damped with the new massive quantity, the Casimir mass~\eq{eq:M:infty}, which is unexpectedly more than three times lighter than the mass of the lowest glueball~\eq{eq:M:glueball} in the model. As the chromometallic wires become less perfect the Casimir mass increases towards the lowest glueball mass.

At small interwire separations the Casimir energy gets a small anomalous scaling dimension~\eq{eq:nu:infty} so that the short-distance Casimir interaction is slightly different from the canonical three-level $R^{-2}$ behavior~\eq{eq:V:Cas:tree}.

Finally, we observed that the Casimir effect induces a (smooth) confinement-deconfinement transition of the gluonic fields in between the wires: the expectation value of the Polyakov line in the space in between the wires increases as the wires get closer. The relatively low value of the Casimir mass, discussed earlier, may be a result the gradually induced deconfinement in a shrinking finite geometry which weakens the mass gap generation of the zero-temperature Yang-Mills theory.

\acknowledgments

The numerical simulations were performed at the computing cluster Vostok-1 of Far Eastern Federal University. The research was carried out within the state assignment of the Ministry of Science and Education of Russia (Grant No. 3.6261.2017/8.9).

\end{document}